\documentclass[letter]{aa} % for the letters

\usepackage{graphicx}
\usepackage{txfonts}
\usepackage{amstext}
\usepackage[colorlinks=true,allcolors=blue]{hyperref}

\usepackage{hyperref}
\hypersetup{colorlinks = true, citecolor = {blue}}

\newcommand{\vr}{{\mathbf r}}
\newcommand{\vv}{{\mathbf v}}

\def\PGRAPE{$\varphi-$GRAPE}

\def\gapprox{\;\rlap{\lower 3.0pt                       % approximately smaller
        \hbox{$\sim$}}\raise 2.5pt\hbox{$>$}\;}
\def\lapprox{\;\rlap{\lower 3.1pt                       % approximately smaller
        \hbox{$\sim$}}\raise 2.7pt\hbox{$<$}\;}

\newcommand{\be}{ \begin{equation} }
\newcommand{\ee}{\end{equation}}

\newcommand{\ben}{\begin{enumerate}}
\newcommand{\een}{\end{enumerate}}

\begin{document}
   
\title{Collinder 135 and UBC 7: A Physical Pair of Open Clusters}
%%{Are Collinder 135 and UBC 7 born together?}
%%%\title{Do Cr~135 and UBC 7 open clusters form a physical pair?}
\titlerunning{Cr~135 and UBC 7: A Pair of Clusters}

%%{Are Cr~135 and UBC 7 born together?}
%\subtitle{Do Cr~135 and UBC 7 open clusters form a physical pair?}
\author{Dana~A.~Kovaleva
\inst{1}\fnmsep\thanks{\tt dana@inasan.ru}
\and
Marina~Ishchenko
\inst{2}
%%%\fnmsep\thanks{\tt https://orcid.org/0000-0002-6961-8170}
\and
Ekaterina~Postnikova
\inst{1}%\fnmsep\thanks{}
\and
Peter~Berczik
\inst{3,4,2}
%%%\fnmsep\thanks{\tt https://orcid.org/0000-0003-4176-152X}
\and
Anatoly E. Piskunov
\inst{1}%\fnmsep\thanks{}
\and
Nina~V.~Kharchenko
\inst{1,2}%\fnmsep\thanks{}
\and
Evgeny Polyachenko
\inst{1}%\fnmsep\thanks{}
\and
Sabine~Reffert
\inst{5}%\fnmsep\thanks{}
\and
Kseniia~Sysoliatina
\inst{4}%\fnmsep\thanks{}
\and
Andreas~Just
\inst{4}%\fnmsep\thanks{}
}

\institute{Institute of Astronomy, Russian Academy of Sciences, 48 Pyatnitskya St, Moscow 119017, Russia			
           \and
           Main Astronomical Observatory, National Academy of Sciences of Ukraine, 27 Akademika Zabolotnoho St, 03143 Kyiv, Ukraine   
           %\email{}
           %\thanks{}
           \and
           National Astronomical Observatories and Key Laboratory of Computational Astrophysics, 
		   Chinese Academy of Sciences, 20A Datun Rd., Chaoyang District, Beijing 100101, China
           \and
           Zentrum f\"ur Astronomie der Universit\"at Heidelberg, Astronomisches Rechen-Institut, M\"onchhofstr 12-14, 69120 Heidelberg, Germany
           \and
           Landessternwarte, Zentrum f\"ur Astronomie der Universit\"at Heidelberg, K\"onigstuhl 12, 69117, Heidelberg, Germany
           }
   
\date{Received 19 Aug 2020 / Accepted 4 September 2020}

\abstract    
% {} leave it empty if necessarys
{Given the closeness of the two open clusters Cr 135 and UBC 7 on the sky, we investigate the possibility of the two clusters to be physically related.}
{We aim to recover the present-day stellar membership in the open clusters Collinder~135 and UBC~7 (300~pc from the Sun), to constrain their kinematic parameters, ages and masses, and to restore their primordial phase space configuration.}
{The most reliable cluster members are selected with our traditional method modified for the use of Gaia DR2 data. Numerical simulations use the integration of cluster trajectories backwards in time with our original high order Hermite4 code \PGRAPE.}
{We constrain the age, spatial coordinates and velocities, radii and masses of the clusters. We estimate the actual separation of the cluster centres equal to 24 pc. The orbital integration shows that the clusters were much closer in the past if their current line-of-sight velocities are very similar and the total mass is more than 7 times larger the mass of the determined most reliable members.}
{We conclude that the two clusters Cr 135 and UBC 7 might very well have formed a physial pair, based on the observational evidence as well as numerical simulations. The probability of a chance coincidence is only about $2\%$.}

\keywords{open clusters and associations: individual -- numerical integration in Galactic potential: star cluster orbits -- binary star cluster evolution: initial conditions for star clusters}
     
\maketitle

%%%%%%%%%%%%%%%%%%%%%%%%%%%%%%%%%%%%%%%%%%%%%%%%%%%%%%%%%%%%%%%%%%%%
\section{Introduction}% -- on Dana++ 
\label{sec:Intr}
%%%%%%%%%%%%%%%%%%%%%%%%%%%%%%%%%%%%%%%%%%%%%%%%%%%%%%%%%%%%%%%%%%%%

Gravitating matter tends to clusterise and form objects on different scales. It is suspected that, similar to stars, star clusters may form groups  \citep{1996ApJ...466..802E, 1997AJ....113..249F, 2010MNRAS.403..996B, 2013MNRAS.434..313G} and physical pairs. The fraction of these {\it binary clusters} was first estimated at a level of 20\,\%~\citep{1976Ap.....12..204R} but later decreased to 10\,\%~\citep{2009A&A...500L..13D}. The latter and some other works~\citep[see, e.g.,][]{2010A&A...511A..38V, 2017A&A...600A.106C} give lists of potential cluster binaries and even triples, however many of them were eventually dismissed~\citep{2010A&A...511A..38V, 2018A&A...619A.155S}.

Using the Gaia DR2 catalogue \cite{2018A&A...619A.155S} have shortened the list to eleven pairs. The prime candidates in our Galaxy are $h$ and $\chi$~Per -- large and massive open clusters, close to each other on the sky, with nearly the same distance of $2.2\pm0.2$~kpc from the Sun \citep[e.g.,][]{2019A&A...624A..34Z}. This large distance, however, makes their detailed study difficult.

Meanwhile, the Magellanic Clouds may contain a lot of binary clusters~\citep[e.g.,][]{1988MNRAS.230..215B, 1990A&A...230...11H, 1999AcA....49..165P, 2002A&A...391..547D}. Numerical simulations  \citep[e.g.,][] {2007MNRAS.374..931P} and spectroscopic investigations \citep[e.g.,][]{2019A&A...622A..65M} prove the existence of physical pairs of clusters in the LMC. The difference in the number of binary clusters between the Magellanic Clouds and our own Galaxy can be explained either by some kind of observational bias \citep[e.g.][]{2010A&A...511A..38V} or by peculiarities of the formation and/or destruction processes in different types of galaxies. 

Numerical simulations of a test binary cluster in the Galactic tidal field demonstrate a complicated dependence between its initial properties and future history before the components actually merge and produce a single rotating star cluster \citep{2016MNRAS.457.1339P}. These results were not tested on real Galactic clusters because of the lack of relevant observational data.

The Gaia mission~\citep{2018A&A...616A...1G} resulted in the discovery of many previously unknown clusters \citep[e.g.][]{2018A&A...618A..59C, 2018A&A...618A..93C, 2020A&A...635A..45C}. One of them, UBC~7, was found at a distance of $\sim 300$~pc from the Sun near the well-known open cluster Collinder~135 (hereafter Cr~135) and mentioned as probably related to it \citep{2018A&A...618A..59C}. Before Gaia, the stars now attributed to UBC~7 were considered as part of Cr~135.
With the use of Gaia DR2, in this paper, we first aim at disentangling the stellar membership between the two clusters and at obtaining the most probable kinematic parameters of the clusters in 6D space. We use these data to recover plausible initial conditions of the clusters,
enabling a future study of their detailed evolution with full-scale N-body simulations.

%%%%%%%%%%%%%%%%%%%%%%%%%%%%%%%%%%%%%%%%%%%%%%%%%%%%%%%%%%%%%%%%%%%%%
\section{Characterising Cr~135 and UBC~7 with Gaia DR2}
\label{sec:data}
%%%%%%%%%%%%%%%%%%%%%%%%%%%%%%%%%%%%%%%%%%%%%%%%%%%%%%%%%%%%%%%%%%%%%

Cr~135 and UBC~7 are located at a distance of approximately 300~pc in the Vela-Puppis star formation region ($245\,{\rm deg} \lesssim l \lesssim 265$ deg, $-15\,{\rm deg} \lesssim b \lesssim -5$ deg), which was recently thoroughly investigated with respect to its large-scale structure and kinematics using Gaia DR2 data \citep{2019A&A...626A..17C, 2019A&A...621A.115C, 2020MNRAS.491.2205B}. 
In particular, the region at distances between 250 and 500 pc~hosts several young open clusters divided into groups of similar age. The oldest group \citep[30 to 50 Myr according to different authors, e.g.][]{2019A&A...626A..17C,  2020MNRAS.491.2205B} includes Cr~135 and UBC~7. 

%--------------------------------------------------------------------
\subsection{Cluster membership}
\label{sec:MWSC}

%--------------------------------------------------------------------

As a compromise allowing to study the outer regions of the clusters and simultaneously avoid contamination from nearby groups, we
used all sources of Gaia DR2 within a radius of $6.5$ deg around the center of Cr~135 ($\alpha=108.3$\,deg, $\delta=-37.35$\,deg, also applicable to UBC~7) that satisfy requirements for ``astrometrically pure'' solutions according to \cite{2018A&A...616A...2L} and technical note GAIA-C3-TN-LU-LL-124-01\,\footnote{\url{http://www.rssd.esa.int/doc\_fetch.php?id=3757412}}.
This includes requirements $ruwe<1.4$, using limits for the flux excess factor \citep{2018A&A...616A...2L}, and selecting sources with $\sigma_\varpi / \varpi \leq 10\%$. The number of sources satisfying these requirements is 411,153. For each of these sources, we calculate the cluster membership probability (MP) for Cr~135 or UBC~7 following the principles formulated in \cite{2012A&A...543A.156K} with specific adjustments to use Gaia DR2 data described below. 

Initial estimates of basic parameters $\overline{\mu}^k_l$, $\overline{\mu}^k_b$, $\overline{\varpi}^k$, age $T^k$, $E^k({\rm BP-RP})$  of Cr~135 ($k=1$) and UBC~7 ($k=2$) are obtained for a subsample of evident members of the two clusters based on a visual analysis of astrometric and photometric diagrams: the vector point diagram (VPD), a parallax vs.\ magnitude plot ($\varpi,\,G$), and a Gaia colour--magnitude diagram (CMD). The parameters are further adjusted along with a list of cluster members in an iterative procedure \citep[see details in][] {2012A&A...543A.156K}. 
Isochrones for Gaia DR2 passbands from \cite{2018A&A...619A.180M} are obtained from the Padova webserver CMD3.3\,\footnote{\url{http://stev.oapd.inaf.it/cmd}}, based on the calculations by \cite{2012MNRAS.427..127B} for solar metallicity $Z=0.0152$. We apply systematic corrections for the $G$ magnitudes of sources\footnote{\url{https://www.cosmos.esa.int/web/gaia/dr2-known-issues\#PhotometrySystematicEffectsAndResponseCurves}} to adjust for the use of these passbands. 
We use a relation between $E({\rm BP-RP})$ and $A_{G}$ based on coefficients provided at CMD3.3 for $A_\lambda/A_V$ for Gaia photometric bands following the relations by \cite{1989ApJ...345..245C} and \citet{1994ApJ...422..158O},
which leads to $A_{G}/E({\rm BP-RP}) \approx 2.05$.

We find that the ages of the two clusters cannot be distinguished, and neither their reddenings, so we use only one value for both clusters. In contrast, proper motions and mean parallaxes of the selected groups of stars are clearly different. 

The modification to derive the mean parameters of the clusters with respect to that described in \cite{2012A&A...543A.156K} involves taking into account the parallax probability $P^{i,k}_{\rm \varpi}$ and the photometric probability $P^{i,k}_{\rm ph}$ based on the $(G, {\rm BP-RP})$ CMD to derive MP. The resulting MP is given by:
\begin{equation}
    P^{i,k}=\min(P^{i,k}_{\rm kin}, P^{i,k}_{\rm \varpi}, P^{i,k}_{\rm ph}).
\end{equation}
The values for parameters $\varepsilon_{\mu_l}, \varepsilon_{\mu_b}, \varepsilon^i_{\rm \varpi}, \varepsilon^i_{\rm ph}$, characterising the expected dispersion of cluster member parameters in calculation of the probabilities $P^{i,k}_{\rm kin}$, $P^{i,k}_{\rm \varpi}$, and  $P^{i,k}_{\rm ph}$, are estimated as follows. 
The dispersion of proper motions in the clusters at 300~pc distance is mainly due to actual velocity dispersion of cluster members rather than due to proper motion errors of individual stars (mean value of 0.15~mas/yr, maximum of 0.45~mas/yr for cluster member candidates). We set $\varepsilon_{\mu_l}, \varepsilon_{\mu_b}= 1.8$~mas/yr.

In turn, the dispersion of parallaxes is dominated by the accuracy of the observations rather than by the actual dispersion of the distances. The pre-defined limit of a relative error of $10\%$ corresponds to approximately 25~pc at the given distance, which is definitely more than the spatial dispersion of members of moderate-size open clusters. For the allowances for parallaxes of possible cluster members, we use an expression for the external calibration of the parallax error\,\footnote{\label{Lindegren}\url{https://www.cosmos.esa.int/documents/29201/1770596/Lindegren\_GaiaDR2\_Astrometry\_extended.pdf/1ebddb25-f010-6437-cb14-0e360e2d9f09}}. The value of $\varepsilon^i_{\rm ph}$ is defined from the dispersion of mean parallaxes obtained at the initial selection of reliable cluster members, equal to $0.12$~mas for each cluster, and individual photometric error $\sigma^i_G$ estimated by flux and flux error in G.

\begin{figure*}[!htb]
 \centering
\includegraphics[width=0.9\textwidth]{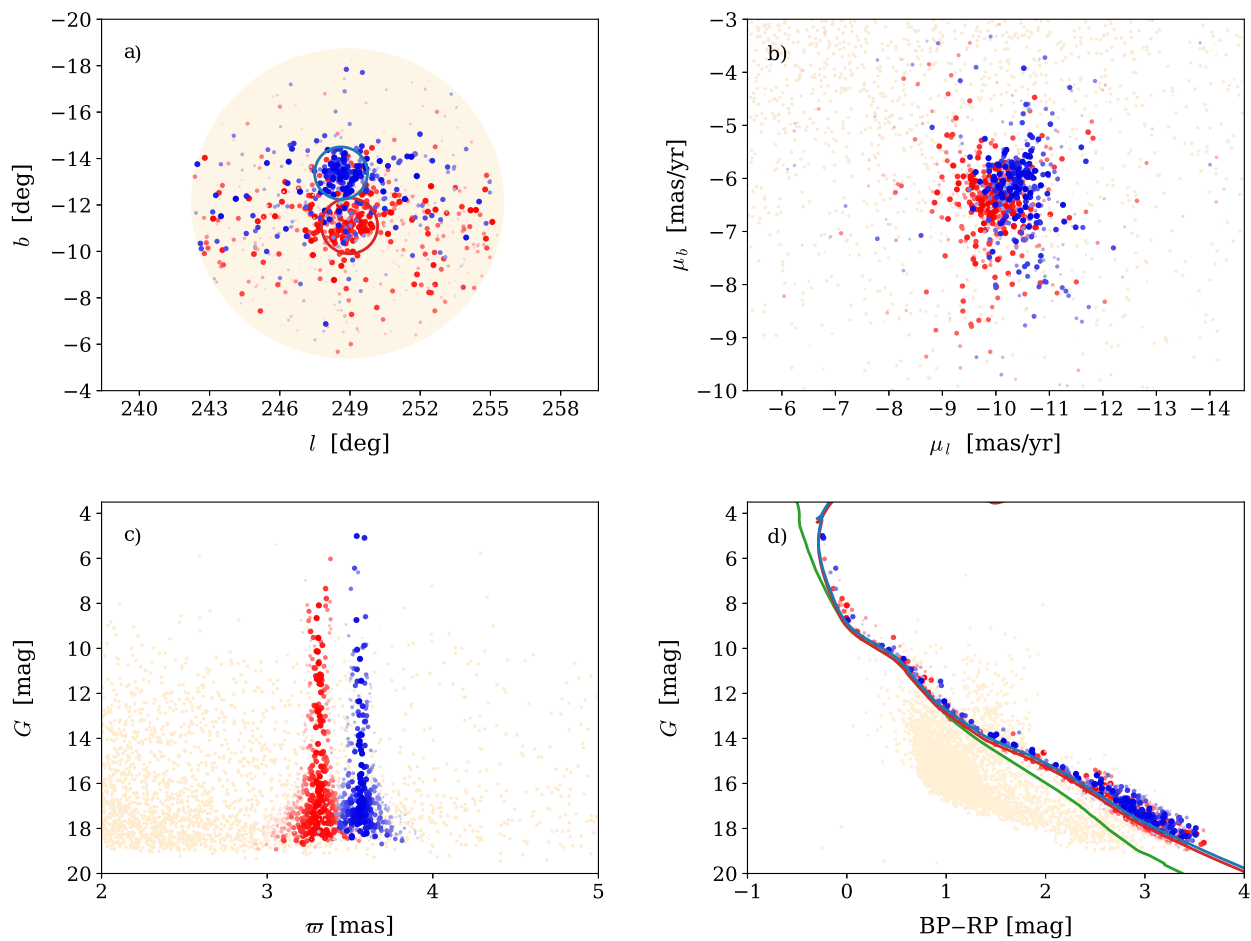}
  \caption{Observational data for Cr~135 and UBC~7: Red dots represent sources having larger Cr~135 MPs, and blue dots sources having larger UBC~7 MPs. Shade dots are for Gaia DR2 sources having  $P^{i,k}<0.01$ for both clusters. (a) Location of probable members of Cr~135 and UBC~7 in the Galactic $l,b$-plane; (b) VPD; (c)~magnitude -- parallax diagram; (d) CMD. In (d) red and blue lines represent isochrones for 40~Myr. The green line is for zero-age main sequence built in the present study as hot envelope of the related set of Padova isochrones of different ages. Absolute magnitudes and colours are reduced to the apparent scale using mean parallaxes and reddening of central probable members. Large circles in (a) mark central regions of the clusters (see Sec.\ref{sec:param}). }
  \label{fig:obsdata}
\end{figure*}

Fig.~\ref{fig:obsdata} represents the distribution of the selected members over the  map of the considered region (a), VPD (b), parallax-magnitude diagram (c), and CMD (d). One can see that the dispersion of observational data of probable members allows to distinguish the clusters in panels (a,b,c) but not in (d).

%--------------------------------------------------------------------
\subsection{Parameters of the clusters}
\label{sec:param}

With the MPs to the two clusters figured out, we derive ages between 40 and 50 Myr. We select sources with $P^{i,k}>0.6$ as the most probable members of Cr~135 and UBC~7. 244 and 184 stars are identified as probable members of Cr~135 and  of UBC~7, respectively. Twelve stars are probable members of both clusters.

The distribution of probable cluster members in the sky looks like two relatively compact cores surrounded by a halo of stars extending to more than 5~deg from the centres of the clusters. We estimate the probability of a random contamination of the dataset with field stars, applying the search procedure to stars satisfying the same conditions as those selected as probable members of Cr~135 and UBC~7 in eight equal areas at the perimeter of the Vela-Puppis region. The resulting selections provide 0 to 7 false members in the area. Thus, the extended structures hosting tenths of probable members around Cr~135 and UBC~7 are not due to occasional field contamination and may be a part of the extended common halo of the two clusters, or a signature of filamentary structures discovered by \cite{2020MNRAS.491.2205B}. 
Sometimes, it is difficult to attribute a star satisfying the membership conditions to one cluster or the other, or to the outer structure. 
We select only the probable members belonging to the central parts of the two clusters to access motion of the cluster centres.

The coordinates and parallaxes of the cluster centres are derived from the position of the maxima of the distribution of stellar density of probable members of Cr~135 and UBC~7, respectively. The radii of the central parts of clusters are selected as estimates of their apparent half-mass radii \citep[HMR, see, e.g., ][]{2011A&A...531A..92R}. For this purpose, masses are attributed to probable members by setting them onto the isochrone for 40~Myr (unresolved binaries neglected). Such lower boundary estimates for the total mass based on the most probable members with pure astrometric data are $126 M_\odot$ and $87 M_\odot$ for Cr~135 and UBC~7, respectively. The consequent angular HMR are $1.20$~deg and $1.13$~deg. These estimates agree with those for tidal radii according to King's formula \citep{1962AJ.....67..471K}, and half-number radii, within $10\%$ to $30\%$. Further, we use the HMR to define the central parts of the clusters (shown with large circles in Fig.~\ref{fig:obsdata}\,a) to obtain estimates for the mean basic parameters of Cr~135 and UBC~7. The central parts contain 91 stars for Cr~135 and 80 stars for UBC~7.

In Table~\ref{tab:data}, we quote cluster parameters computed via averaging of individual data on $l,b,\varpi, \mu_l, \mu_b$ for probable members residing within the central areas. For $l, b$ the accuracies are computed as standard deviations. For $\varpi, \mu_l, \mu_b$ the accuracies are computed as a combination of error of the mean and expected systematic error estimate\,\textsuperscript{\ref{Lindegren}}. The number of stars with radial (line-of-sight, LOS) velocity measurements in the central parts of clusters is low, 14 and 5 sources only for Cr~135 and UBC~7. To improve the statistics, we take into account not only astrometrically pure but all probable cluster members with LOS velocity measurments (30 and 24 sources, respectively), and quote their median values. The accuracies are evaluated as $(V_r^{Q3}-V_r^{Q1})/2$ where $V_r^{Q1},V_r^{Q3}$ are LOS velocities corresponding to lower and upper quartiles.
\begin{table*}[tbp]
\setlength{\tabcolsep}{4pt}
\centering
\caption{The cluster main parameters evaluated for the most reliable members.}
\label{tab:data}
\begin{tabular}{ccccccccccc}
\hline 
\hline 
Cluster & $N$ & mass, $M_\odot$ & $N_c$ & $l$, deg & $b$, deg & $\varpi$, mas & $\mu_l$, mas/yr & $\mu_b$, mas/yr & $N_{V_r}$ & $V_r$, km/s \\
\hline
Cr~135 & 244 & 126 & 91 & 248.98 $\pm$ 0.06 & $-$11.10 $\pm$ 0.05 & 3.31 $\pm$ 0.02 & $-$9.92 $\pm$ 0.05 & $-$6.47 $\pm$0.06 & 30 & 17.4 $\pm$ 1.3 \\[-10pt]
UBC~7 & 184 & 87 & 80 & 248.62 $\pm$ 0.04& $-$13.37 $\pm$ 0.05 & 3.56 $\pm$ 0.02 &$-$10.25 $\pm$ 0.05 &$-$5.98 $\pm$ 0.05 & 24 & 16.7 $\pm$ 1.5\\
\hline 
\end{tabular}
\vspace{6pt}
\end{table*}

The kinematics and age obtained for the two considered clusters are very similar. The separation between the centres of Cr~135 and UBC~7 is
$24.2 \pm 2.1$~pc, the difference in proper motion
$0.6 \pm 0.1$~mas/yr (the relative tangential velocity is $1.42 \pm 0.15$~km/s).
A simple model of the local vicinity based on parameters by \cite{2019A&A...626A..17C} 
hosts six clusters in a region with dimensions [100, 50, 200]~pc with $\mu_l, \mu_b$ range within 6~mas/yr. If their positions and proper motions are distributed randomly, 
the likelihood of some pair of clusters simultaneously having spatial distance less than 25~pc and differences in proper motion less than 0.7~mas/yr is quite small ($P_r=2.4\%$), so that it seems unlikely that Cr~135 and UBC~7 are located so close together by chance.

%--------------------------------------------------------------------

%%%%%%%%%%%%%%%%%%%%%%%%%%%%%%%%%%%%%%%%%%%%%%%%%%%%%%%%%%%%%%%%%%%%%

\section{Orbital integration of the star clusters in the Milky Way potential}
\label{sec:dyn}

In this paper, we restrict ourselves to the simplest model of the star clusters as attracting point masses orbiting in the fixed Milky Way external potential~\citep{2011A&A...536A..64E}. We carry out the integration backwards in time up to $-50$\,Myr
using our own developed high order Hermite4 code \PGRAPE~ \citep{HGM2007}. 
The current version of the \PGRAPE\footnote{\url{ftp://ftp.mao.kiev.ua/pub/berczik/phi-GRAPE/}} 
code uses GPU/CUDA based GRAPE emulation YEBISU library \citep{NM2008}; it was tested and successfully applied in our previous large scale simulations~\citep{2020MNRAS.492.4819P, 2016MNRAS.460..240K, 2014ApJ...780..164W, 2014ApJ...792..137Z, 2012ApJ...748...65L, 2012ApJ...758...51J}. 

The largest uncertainty of kinematic data resides in the LOS velocities. So we probe these velocities on a uniform mesh of $101\times101$ runs covering $\pm3\sigma$-confidence intervals, while other parameters are kept fixed at the averaged values. A pair of initial ($t=0$) LOS velocities gives coordinates $\vr_k(t)$ and velocities $\vv_k(t)$ of the clusters. Our fiducial series of runs assumes the cluster mass ratio close to that from Table\,\ref{tab:data}, but the actual values $M_1 = 465\,M_\odot$ and $M_2 = 302\,M_\odot$ are larger because the table data contain the present-day masses of the most reliable members only. 

\begin{figure}%[!htb]
    \centering
    \includegraphics[width=\columnwidth]{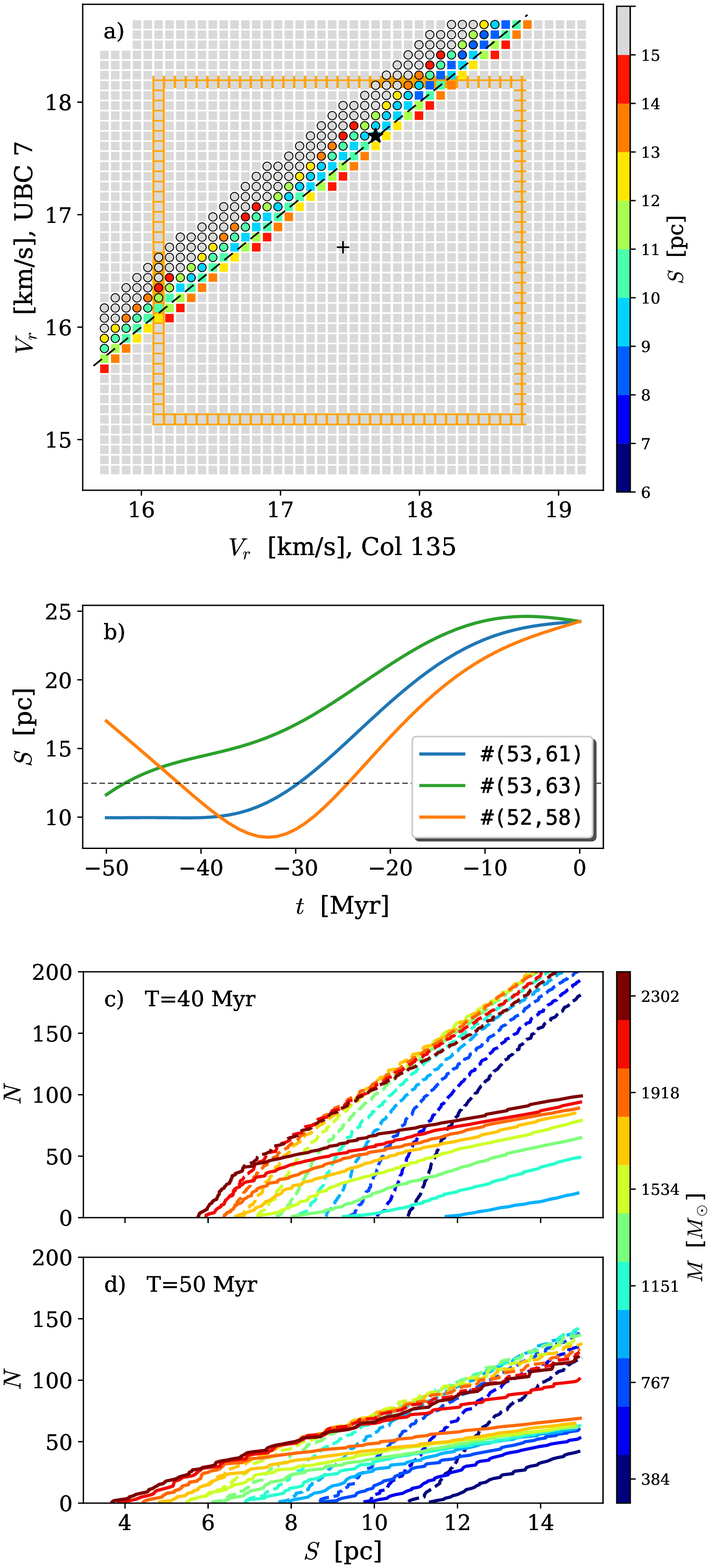}
    \caption{Numerical backward integration of two point masses representing the clusters in the external Milky Way potential. \newline
        (a) The run mesh (fragment) for cluster age $T=50$ Myr: small squares represent individual runs with different present-day ($t=0$) LOS velocities. Coloured squares mark runs with primordial separations smaller than 15 pc. The black plus and star denote the central $V_r$ values and run {\tt \#(53,61)} given in panel (b). Rimmed circles mark runs with negative primordial energy $E_{\rm K}$. The dashed black line and the orange rectangle depict the line of equal LOS velocities and the runs within $\pm 1\sigma$ from mid values. \newline
        (b) Typical curves of separation $S(t)$. 
        \newline
        (c, d) The cumulative number of runs vs. the primordial separation below 15 pc for 40 and 50\,Myr. The dashed lines show all runs within $1\sigma$-rectangle, the solid lines show only runs with negative energy $E_{\rm K}$ (bound). Colours of the lines code the total mass $M$ of the clusters.
    }
    \label{fig:sim}
\end{figure}

First of all, we are looking for runs with small primordial separations $S(-T)$, where $S(t) = |\vr_1(t) - \vr_2(t)|$ is the distance between the clusters. Small colour squares in Fig.\,\ref{fig:sim}\,a) mark runs with separations $S$ below 15 pc at age $T=50$ Myr (for $T=40$ Myr the plot is similar). They settle in a narrow band near the lines of equal LOS velocities. The rimmed circles mark `bound' runs with negative (specific) energy
\be 
  {E}_{\rm K} \equiv \frac{|\vv_1-\vv_2|^2}2 - \frac{GM}{S}\,,\quad M \equiv M_1+M_2\,
\ee
that does not account for the external Galactic field. At age $40$\,Myr, none of these runs are bound, although some with the positive $E_{\rm K}$ lead to primordial separations below 10 pc. At 50\,Myr age the runs show a minimum separation of 8.8 pc, and for the bound runs 9.1 pc.  

Fig.\,\ref{fig:sim}\,b) presents typical separation curves. Runs in the grey zone and some on both sides of the colour zone have curves similar to {\tt \#(52,58)} with a trend of initial convergence and then increase of the separation. Monotonic separation growth is met in runs with very similar LOS velocities (either with negative and positive primordial energy), see, e.g., runs {\tt \#(53,61)} and {\tt \#(53,63)}. The dashed line indicate the position of the tidal (Jacobi) radius
for the system of two attracting points.
Clusters formed closer than the Jacobi radius with the total energy below the Jacobi energy will stay close forever. In contrast, clusters formed with energy larger than the Jacobi energy can defy mutual attraction either because of the tidal force ($E_{\rm K}<0$) or high relative velocity ($E_{\rm K}>0$).

To account for the mass uncertainty, in addition to the fiducial series we explored ten more series increasing and decreasing masses proportionally in the range of $M$ between 380 and 2300 $M_\odot$. Similarly to the fiducial series, the runs with small separations always settle near the lines of equal LOS velocities.

Fig.\,\ref{fig:sim}\,c),\,d) present the cumulative number of runs within the 1$\sigma$-rectangle vs. the primordial separation at 40 and 50~Myr, respectively. The fiducial and lighter series give no bound runs at 40~Myr, although there are some runs with $E_{\rm K}>0$. Meanwhile, the 50~Myr plots contain bound runs at any explored value of the total mass $M$. 

At the high-mass end, all runs with primordial separation $S$ smaller than 15~pc are bound at 50~Myr, while only 46\% are bound at 40~Myr. The separation is smaller and the width of the colour bands is narrower in $T=50$ Myr plots. The latter explains smaller cumulative numbers at $S\sim 15$ pc compared to $T=40$ Myr.
If we look at the bound runs only, clusters of age 40~Myr require a total mass $M$ above $\sim 1500$ $M_\odot$ to obtain small primordial separations starting from reasonable values of LOS velocities. The 50~Myr clusters are less restrictive in this aspect. 

%%%%%%%%%%%%%%%%%%%%%%%%%%%%%%%%%%%%%%%%%%%%%%%%%%%%%%%%%%%%%%%%%%%%%

%%%%%%%%%%%%%%%%%%%%%%%%%%%%%%%%%%%%%%%%%%%%%%%%%%%%%%%%%%%%%%%%%%%%%
\section{Conclusions}
\label{sec:conc}

Based on Gaia DR2 data, we have selected probable members of Cr~135 and UBC~7 and determined their parameters in 6D space ($l, b, \varpi, \mu_l, \mu_b, V_r$).
The clusters are proven to be close but distinctly separated, while
their CMDs are indistinguishable. We assume the cluster ages to be virtually equal and estimate it between 40 and 50~Myr. A model with randomized spatial and kinematic parameters shows a likelihood of only $P_r=2.4\%$ for their chance coincidence. Besides, the observations show coronae enveloping both clusters. 
This suggests their possible physical binarity.

The clusters may have formed closer together than they appear now. In order to show this, we use a simple model in which the clusters are replaced by point masses and integrate backwards in time in the fixed external Galactic potential. The masses of the points were constant during the integration, and their values accounted also for the unresolved stars and possible mass loss. 

Given the uncertainty in the observational data, we performed an optimisation over initial LOS velocities leading to desirable runs with small primordial separations. They are obtained only in the case of very similar LOS velocities. Then we explore the question of how many desirable runs with plausible LOS velocities (within 1$\sigma$-confidence rectangle) occur in series with different total mass and age. We report that independent of age, clusters with a total mass $M \gtrsim 1500\,M_\odot$ are favourable for the scenario in which the clusters were initially close in the beginning and then were tidally separated. On the other hand, relatively young clusters with a total mass $M \lesssim 750\,M_\odot$ require LOS velocities above the confidence intervals determined from observations.

The initial masses of the clusters are very uncertain. Accounting for incompleteness and mass loss by stellar evolution may result in a factor of two larger initial masses compared to the observed masses given in this paper, which is insufficient to reach the high mass regime of our investigations. But since clusters are formed in molecular clouds with low star formation efficiency, they are most probably supervirial after gas expulsion. This leads to a significant dynamical mass loss on a dynamical timescale of 10--20\,Myr. In case of a centrally peaked star formation efficiency the mass of the surviving cluster can be as low as 5\% of the initial mass \citep{2017A&A...605A.119S,2018ApJ...863..171S}. The observed extended corona of cluster stars around Cr~135 and UBC~7 is a hint to that kind of strong cluster mass loss by violent relaxation in the first 20\,Myr.

In our simple model for the cluster orbits, the mass loss of the clusters was completely ignored. 
As a continuation of this work, in Paper~II (in preparation) we shall extend our numerical simulations using realistic star cluster N-body modelling 
by forward integrating star-by-star cluster models to the present day and make a direct comparison of the stellar populations to the observations, including selection effects and binary stars.

%%%%%%%%%%%%%%%%%%%%%%%%%%%%%%%%%%%%%%%%%%%%%%%%%%%%%%%%%%%%%%%%%%%%%
\begin{acknowledgements}

This work has made use of data from the European Space Agency (ESA) mission {\it Gaia} (\url{https://www.cosmos.esa.int/gaia}), processed by the {\it Gaia} Data Processing and Analysis Consortium (DPAC, \url{https://www.cosmos.esa.int/web/gaia/dpac/consortium}). Funding for the DPAC has been provided by national institutions, in particular the institutions participating in the {\it Gaia} Multilateral Agreement. The use of TOPCAT, an interactive graphical viewer and editor for tabular data \citep{2005ASPC..347...29T}, is acknowledged.

%%% Moscow %%%
The reported study was partly funded by RFBR and DFG according to the research project No. 20-52-12009.
   
%%% Peter & Marina %%%
The work of PB and MI was supported by the Deutsche Forschungsgemeinschaft (DFG, German Research Foundation) Project-ID 138713538, SFB 881 ("The Milky Way System") and by the Volkswagen Foundation under the Trilateral Partnerships grant No. 97778. 

PB acknowledges support by the Chinese Academy of Sciences (CAS) through the Silk Road Project at NAOC, the President’s International Fellowship (PIFI) for Visiting Scientists program of CAS and the National Science Foundation of China (NSFC) under grant No. 11673032. 

MI acknowledges support by the National Academy of Sciences of Ukraine under the Young Scientists Grant No. 0119U102399. 

The work of PB was also partially supported under the special program of the National Academy of Sciences of Ukraine "Support for the development of priority fields of scientific research" (CPCEL 6541230). 

%%% Peter & Marina %%%

We thank the referee for the helpful comments. 
\end{acknowledgements}

%%%%%%%%%%%%%%%%%%%%%%%%%%%%%%%%%%%%%%%%%%%%%%%%%%%%%%%%%%%%%%%%%%%%%

%%%%%%%%%%%%%%%%%%%%%%%%%%%%%%%%%%%%%%%%%%%%%%%%%%%%%%%%%%%%%%%%%%%%%

\normalfont
\bibliographystyle{aa}  % style aa.bst
\bibliography{main}   % your references Yourfile.bib

\begin{thebibliography}{44}
\expandafter\ifx\csname natexlab\endcsname\relax\def\natexlab#1{#1}\fi

\bibitem[{{Beccari} {et~al.}(2020){Beccari}, {Boffin}, \&
  {Jerabkova}}]{2020MNRAS.491.2205B}
{Beccari}, G., {Boffin}, H. M.~J., \& {Jerabkova}, T. 2020, \mnras, 491, 2205

\bibitem[{{Bhatia} \& {Hatzidimitriou}(1988)}]{1988MNRAS.230..215B}
{Bhatia}, R.~K. \& {Hatzidimitriou}, D. 1988, \mnras, 230, 215

\bibitem[{{Bonatto} \& {Bica}(2010)}]{2010MNRAS.403..996B}
{Bonatto}, C. \& {Bica}, E. 2010, \mnras, 403, 996

\bibitem[{{Bressan} {et~al.}(2012){Bressan}, {Marigo}, {Girardi}, {Salasnich},
  {Dal Cero}, {Rubele}, \& {Nanni}}]{2012MNRAS.427..127B}
{Bressan}, A., {Marigo}, P., {Girardi}, L., {et~al.} 2012, \mnras, 427, 127

\bibitem[{{Cantat-Gaudin} {et~al.}(2018){Cantat-Gaudin}, {Jordi}, {Vallenari},
  {Bragaglia}, {Balaguer-N{\'u}{\~n}ez}, {Soubiran}, {Bossini}, {Moitinho},
  {Castro-Ginard}, {Krone-Martins}, {Casamiquela}, {Sordo}, \&
  {Carrera}}]{2018A&A...618A..93C}
{Cantat-Gaudin}, T., {Jordi}, C., {Vallenari}, A., {et~al.} 2018, \aap, 618,
  A93

\bibitem[{{Cantat-Gaudin} {et~al.}(2019{\natexlab{a}}){Cantat-Gaudin}, {Jordi},
  {Wright}, {Armstrong}, {Vallenari}, {Balaguer-N{\'u}{\~n}ez}, {Ramos},
  {Bossini}, {Padoan}, {Pelkonen}, {Mapelli}, \&
  {Jeffries}}]{2019A&A...626A..17C}
{Cantat-Gaudin}, T., {Jordi}, C., {Wright}, N.~J., {et~al.} 2019{\natexlab{a}},
  \aap, 626, A17

\bibitem[{{Cantat-Gaudin} {et~al.}(2019{\natexlab{b}}){Cantat-Gaudin},
  {Mapelli}, {Balaguer-N{\'u}{\~n}ez}, {Jordi}, {Sacco}, \&
  {Vallenari}}]{2019A&A...621A.115C}
{Cantat-Gaudin}, T., {Mapelli}, M., {Balaguer-N{\'u}{\~n}ez}, L., {et~al.}
  2019{\natexlab{b}}, \aap, 621, A115

\bibitem[{{Cardelli} {et~al.}(1989){Cardelli}, {Clayton}, \&
  {Mathis}}]{1989ApJ...345..245C}
{Cardelli}, J.~A., {Clayton}, G.~C., \& {Mathis}, J.~S. 1989, \apj, 345, 245

\bibitem[{{Castro-Ginard} {et~al.}(2020){Castro-Ginard}, {Jordi}, {Luri},
  {{\'A}lvarez Cid-Fuentes}, {Casamiquela}, {Anders}, {Cantat-Gaudin},
  {Mongui{\'o}}, {Balaguer-N{\'u}{\~n}ez}, {Sol{\`a}}, \&
  {Badia}}]{2020A&A...635A..45C}
{Castro-Ginard}, A., {Jordi}, C., {Luri}, X., {et~al.} 2020, \aap, 635, A45

\bibitem[{{Castro-Ginard} {et~al.}(2018){Castro-Ginard}, {Jordi}, {Luri},
  {Julbe}, {Morvan}, {Balaguer-N{\'u}{\~n}ez}, \&
  {Cantat-Gaudin}}]{2018A&A...618A..59C}
{Castro-Ginard}, A., {Jordi}, C., {Luri}, X., {et~al.} 2018, \aap, 618, A59

\bibitem[{{Conrad} {et~al.}(2017){Conrad}, {Scholz}, {Kharchenko}, {Piskunov},
  {R{\"o}ser}, {Schilbach}, {de Jong}, {Schnurr}, {Steinmetz}, {Grebel},
  {Zwitter}, {Bienaym{\'e}}, {Bland -Hawthorn}, {Gibson}, {Gilmore},
  {Kordopatis}, {Kunder}, {Navarro}, {Parker}, {Reid}, {Seabroke}, {Siviero},
  {Watson}, \& {Wyse}}]{2017A&A...600A.106C}
{Conrad}, C., {Scholz}, R.~D., {Kharchenko}, N.~V., {et~al.} 2017, \aap, 600,
  A106

\bibitem[{{de La Fuente Marcos} \& {de La Fuente
  Marcos}(2009)}]{2009A&A...500L..13D}
{de La Fuente Marcos}, R. \& {de La Fuente Marcos}, C. 2009, \aap, 500, L13

\bibitem[{{Dieball} {et~al.}(2002){Dieball}, {M{\"u}ller}, \&
  {Grebel}}]{2002A&A...391..547D}
{Dieball}, A., {M{\"u}ller}, H., \& {Grebel}, E.~K. 2002, \aap, 391, 547

\bibitem[{{Elmegreen} \& {Efremov}(1996)}]{1996ApJ...466..802E}
{Elmegreen}, B.~G. \& {Efremov}, Y.~N. 1996, \apj, 466, 802

\bibitem[{{Ernst} {et~al.}(2011){Ernst}, {Just}, {Berczik}, \&
  {Olczak}}]{2011A&A...536A..64E}
{Ernst}, A., {Just}, A., {Berczik}, P., \& {Olczak}, C. 2011, \aap, 536, A64

\bibitem[{{Fujimoto} \& {Kumai}(1997)}]{1997AJ....113..249F}
{Fujimoto}, M. \& {Kumai}, Y. 1997, \aj, 113, 249

\bibitem[{{Gaia Collaboration} {et~al.}(2018){Gaia Collaboration}, {Brown},
  {Vallenari}, {Prusti}, {de Bruijne}, {Babusiaux}, {Bailer-Jones}, {Biermann},
  {Evans}, \& {Eyer}}]{2018A&A...616A...1G}
{Gaia Collaboration}, {Brown}, A.~G.~A., {Vallenari}, A., {et~al.} 2018, \aap,
  616, A1

\bibitem[{{Gusev} \& {Efremov}(2013)}]{2013MNRAS.434..313G}
{Gusev}, A.~S. \& {Efremov}, Y.~N. 2013, \mnras, 434, 313

\bibitem[{{Harfst} {et~al.}(2007){Harfst}, {Gualandris}, {Merritt}, {Spurzem},
  {Portegies Zwart}, \& {Berczik}}]{HGM2007}
{Harfst}, S., {Gualandris}, A., {Merritt}, D., {et~al.} 2007, \na, 12, 357

\bibitem[{{Hatzidimitriou} \& {Bhatia}(1990)}]{1990A&A...230...11H}
{Hatzidimitriou}, D. \& {Bhatia}, R.~K. 1990, \aap, 230, 11

\bibitem[{{Just} {et~al.}(2012){Just}, {Yurin}, {Makukov}, {Berczik}, {Omarov},
  {Spurzem}, \& {Vilkoviskij}}]{2012ApJ...758...51J}
{Just}, A., {Yurin}, D., {Makukov}, M., {et~al.} 2012, \apj, 758, 51

\bibitem[{{Kennedy} {et~al.}(2016){Kennedy}, {Meiron}, {Shukirgaliyev},
  {Panamarev}, {Berczik}, {Just}, \& {Spurzem}}]{2016MNRAS.460..240K}
{Kennedy}, G.~F., {Meiron}, Y., {Shukirgaliyev}, B., {et~al.} 2016, \mnras,
  460, 240

\bibitem[{{Kharchenko} {et~al.}(2012){Kharchenko}, {Piskunov}, {Schilbach},
  {R{\"o}ser}, \& {Scholz}}]{2012A&A...543A.156K}
{Kharchenko}, N.~V., {Piskunov}, A.~E., {Schilbach}, E., {R{\"o}ser}, S., \&
  {Scholz}, R.-D. 2012, \aap, 543, A156

\bibitem[{{King}(1962)}]{1962AJ.....67..471K}
{King}, I. 1962, \aj, 67, 471

\bibitem[{{Li} {et~al.}(2012){Li}, {Liu}, {Berczik}, {Chen}, \&
  {Spurzem}}]{2012ApJ...748...65L}
{Li}, S., {Liu}, F.~K., {Berczik}, P., {Chen}, X., \& {Spurzem}, R. 2012, \apj,
  748, 65

\bibitem[{{Lindegren} {et~al.}(2018){Lindegren}, {Hern{\'a}ndez}, {Bombrun},
  {Klioner}, {Bastian}, {Ramos-Lerate}, {de Torres}, {Steidelm{\"u}ller},
  {Stephenson}, \& {Hobbs}}]{2018A&A...616A...2L}
{Lindegren}, L., {Hern{\'a}ndez}, J., {Bombrun}, A., {et~al.} 2018, \aap, 616,
  A2

\bibitem[{{Ma{\'\i}z Apell{\'a}niz} \& {Weiler}(2018)}]{2018A&A...619A.180M}
{Ma{\'\i}z Apell{\'a}niz}, J. \& {Weiler}, M. 2018, \aap, 619, A180

\bibitem[{{Mora} {et~al.}(2019){Mora}, {Puzia}, \&
  {Chanam{\'e}}}]{2019A&A...622A..65M}
{Mora}, M.~D., {Puzia}, T.~H., \& {Chanam{\'e}}, J. 2019, \aap, 622, A65

\bibitem[{{Nitadori} \& {Makino}(2008)}]{NM2008}
{Nitadori}, K. \& {Makino}, J. 2008, \na, 13, 498

\bibitem[{{O'Donnell}(1994)}]{1994ApJ...422..158O}
{O'Donnell}, J.~E. 1994, \apj, 422, 158

\bibitem[{{Pietrzynski} \& {Udalski}(1999)}]{1999AcA....49..165P}
{Pietrzynski}, G. \& {Udalski}, A. 1999, \actaa, 49, 165

\bibitem[{{Polyachenko} {et~al.}(2020){Polyachenko}, {Berczik}, {Just}, \&
  {Shukhman}}]{2020MNRAS.492.4819P}
{Polyachenko}, E.~V., {Berczik}, P., {Just}, A., \& {Shukhman}, I.~G. 2020,
  \mnras, 492, 4819

\bibitem[{{Portegies Zwart} \& {Rusli}(2007)}]{2007MNRAS.374..931P}
{Portegies Zwart}, S.~F. \& {Rusli}, S.~P. 2007, \mnras, 374, 931

\bibitem[{{Priyatikanto} {et~al.}(2016){Priyatikanto}, {Kouwenhoven},
  {Arifyanto}, {Wulandari}, \& {Siregar}}]{2016MNRAS.457.1339P}
{Priyatikanto}, R., {Kouwenhoven}, M.~B.~N., {Arifyanto}, M.~I., {Wulandari},
  H.~R.~T., \& {Siregar}, S. 2016, \mnras, 457, 1339

\bibitem[{{R{\"o}ser} {et~al.}(2011){R{\"o}ser}, {Schilbach}, {Piskunov},
  {Kharchenko}, \& {Scholz}}]{2011A&A...531A..92R}
{R{\"o}ser}, S., {Schilbach}, E., {Piskunov}, A.~E., {Kharchenko}, N.~V., \&
  {Scholz}, R.~D. 2011, \aap, 531, A92

\bibitem[{{Rozhavskii} {et~al.}(1976){Rozhavskii}, {Kuz'mina}, \&
  {Vasilevskii}}]{1976Ap.....12..204R}
{Rozhavskii}, F.~G., {Kuz'mina}, V.~A., \& {Vasilevskii}, A.~E. 1976,
  Astrophysics, 12, 204

\bibitem[{{Shukirgaliyev} {et~al.}(2017){Shukirgaliyev}, {Parmentier},
  {Berczik}, \& {Just}}]{2017A&A...605A.119S}
{Shukirgaliyev}, B., {Parmentier}, G., {Berczik}, P., \& {Just}, A. 2017, \aap,
  605, A119

\bibitem[{{Shukirgaliyev} {et~al.}(2018){Shukirgaliyev}, {Parmentier}, {Just},
  \& {Berczik}}]{2018ApJ...863..171S}
{Shukirgaliyev}, B., {Parmentier}, G., {Just}, A., \& {Berczik}, P. 2018, \apj,
  863, 171

\bibitem[{{Soubiran} {et~al.}(2018){Soubiran}, {Cantat-Gaudin},
  {Romero-G{\'o}mez}, {Casamiquela}, {Jordi}, {Vallenari}, {Antoja},
  {Balaguer-N{\'u}{\~n}ez}, {Bossini}, {Bragaglia}, {Carrera}, {Castro-Ginard},
  {Figueras}, {Heiter}, {Katz}, {Krone-Martins}, {Le Campion}, {Moitinho}, \&
  {Sordo}}]{2018A&A...619A.155S}
{Soubiran}, C., {Cantat-Gaudin}, T., {Romero-G{\'o}mez}, M., {et~al.} 2018,
  \aap, 619, A155

\bibitem[{{Taylor}(2005)}]{2005ASPC..347...29T}
{Taylor}, M.~B. 2005, in Astronomical Society of the Pacific Conference Series,
  Vol. 347, Astronomical Data Analysis Software and Systems XIV, ed.
  P.~{Shopbell}, M.~{Britton}, \& R.~{Ebert}, 29

\bibitem[{{V{\'a}zquez} {et~al.}(2010){V{\'a}zquez}, {Moitinho}, {Carraro}, \&
  {Dias}}]{2010A&A...511A..38V}
{V{\'a}zquez}, R.~A., {Moitinho}, A., {Carraro}, G., \& {Dias}, W.~S. 2010,
  \aap, 511, A38

\bibitem[{{Wang} {et~al.}(2014){Wang}, {Berczik}, {Spurzem}, \&
  {Kouwenhoven}}]{2014ApJ...780..164W}
{Wang}, L., {Berczik}, P., {Spurzem}, R., \& {Kouwenhoven}, M.~B.~N. 2014,
  \apj, 780, 164

\bibitem[{{Zhong} {et~al.}(2019){Zhong}, {Chen}, {Kouwenhoven}, {Li}, {Shao},
  \& {Hou}}]{2019A&A...624A..34Z}
{Zhong}, J., {Chen}, L., {Kouwenhoven}, M.~B.~N., {et~al.} 2019, \aap, 624, A34

\bibitem[{{Zhong} {et~al.}(2014){Zhong}, {Berczik}, \&
  {Spurzem}}]{2014ApJ...792..137Z}
{Zhong}, S., {Berczik}, P., \& {Spurzem}, R. 2014, \apj, 792, 137

\end{thebibliography}

%%%%%%%%%%%%%%%%%%%%%%%%%%%%%%%%%%%%%%%%%%%%%%%%%%%%%%%%%%%%%%%%%%%%%

\end{document}